# Very-low temperature synthesis of pure and crystalline lead-free $Ba_{0.85}Ca_{0.15}Zr_{0.1}Ti_{0.9}O_3$ ceramic

Zouhair Hanani[a,b], El-Houssaine Ablouh[c], M'barek Amjoud[a], Daoud Mezzane[a],

Sébastien Fourcade [b], and Mohamed Gouné[b,*]

[a]Laboratory of Condensed Matter and Nanostructures, Cadi Ayyad University, Marrakesh, 40000, Morocco

[b]Institut de Chimie de la Matière Condensée de Bordeaux, University of Bordeaux, Pessac, 33600, France

[c]Laboratory of Bioorganic and Macromolecular Chemistry, Cadi Ayyad University, Marrakesh, 40000, Morocco

[*]Correspondence to: Mohamed.Goune@icmcb.cnrs.fr

**Abstract**

The site-doping strategy of barium titanate ($BaTiO_3$) is a promising way to develop new lead-free materials for energy with enhanced dielectric and piezoelectric properties. A novel strategy to elaborated pure and crystalline $Ba_{0.85}Ca_{0.15}Zr_{0.1}Ti_{0.9}O_3$ (BCZT) ferroelectric powders at low temperature is proposed. It is based on sol-gel method followed by a hydrothermal reaction. The effects of preparation temperature on structure, crystallinity, purity, morphology and particles size distribution of BCZT powders were studied. It was clearly shown that pure and crystalline BCZT could be obtained over 80°C. Furthermore, the BCZT particles exhibit a spherical shape whose mean size increases from 145 nm at 40°C to 160 nm at 80°C. The present study may provide a new strategy to design lead-free ferroelectric materials with enhanced structure and microstructure properties at very low temperature.

## 1. Introduction

Owning to its high ferroelectric, piezoelectric properties, barium titanate ($BaTiO_3$) is regarded as one of potentially promising lead-free materials for developing low-energy consumption deviceslikewireless sensors, actuators and self-powered nanogenerators[1–3]. However, as compared to lead-based perovskite ceramics such as (PZT), pure $BaTiO_3$ exhibits relatively low dielectric constant ($\varepsilon_r$) piezoelectric coefficient ($d_{33}$) [4]. To tackle this problem, researchers are focusing on site-doping strategy to enhance the dielectric and piezoelectric properties of $BaTiO_3$[5,6]. W. Liu and X. Ren[7]have reported non-Pb ceramics of $Ba_{0.85}Ca_{0.15}Zr_{0.1}Ti_{0.9}O_3$with high dielectric constant (18 000) and piezoelectric coefficient (620 pC/N). This promising finding has created a great interest around site-doping, phase's transitions, morphological engineering and sintering processes to develop BCT-BZT ceramic with highly enhanced ferroelectric and piezoelectric properties. Although, the most of these studies have suffered from high-energy consumption, contamination or production scale.

Wet chemical processing like sol-gel, co-precipitation, auto combustion, Pechini method and hydrothermal/solvothermal are extremely effective for the preparation of homogeneous and high purity ferroelectric materials at much lower temperatures compared to traditional ceramic methods.J. P. Praveen et al.[8]haveelaborated $Ba_{0.85}Ca_{0.15}Zr_{0.1}Ti_{0.9}O_3$ceramic by sol-gel processing, and the single-phase perovskite structure appeared at high temperature (700 °C) alongside with some impurity peaks. T.

H. Hsieh et al. [9]have synthetized $Ba_{0.95}Ca_{0.05}Zr_{0.1}Ti_{0.9}O_3$ceramic, using Pechinipolymeric precursor method with a calcination temperature of 700 °C. D. Zhan et al.[10]have reported the elaboration $Ba_{0.95}Ca_{0.05}Zr_{0.3}Ti_{0.7}O_3$ ceramics by a citrate method followed by thermal treatment at 650 °C. Hydrothermal synthesis of nanostructures have attracted much attention for the elaboration of supported or free-standing ferroelectric ceramic with highly improved properties[2,11]. This low cost and eco-friendly techniquecan provide an efficient design of ferroelectric nanomaterials with specifically tailored architectures with an appropriate size and shape[2,12–17]. Few works have reported the preparation of $Ba_{0.85}Ca_{0.15}Zr_{0.1}Ti_{0.9}O_3$ceramic through hydrothermal techniques. S. Hunpratub et al. [18]havestudied the preparation of $Ba_{0.85}Ca_{0.15}Zr_{0.1}Ti_{0.9}O_3$, by hydrothermal processing at 240 °C/16h under different alkaline mediums.They have obtained pseudo-cubic BCZT samples with some crystalline impurities like $BaZrO_3$ and $CaTiO_3$.

At the best of our knowledge, the preparation of pure and crystalline lead-free $Ba_{0.85}Ca_{0.15}Zr_{0.1}Ti_{0.9}O_3$ powders at very low temperatureswas never reported. The results presented in this paper concern the development of a new way for the synthesis of pure and crystalline BCZT lead-free ferroelectric powders at very low temperature (80 °C). The effects of the preparation temperature on the structure, crystallinity, purity, morphology and particles size distribution of BCZT powders were studied.

**2. Experimental**

*2.1. BCZT Powders synthesis*

A new strategy was designed for the elaboration of pure and crystalline BCZT powders through sol-gel followed by a single-step hydrothermal synthesis in high alkaline medium. Under inert atmosphere ($N_2$), appropriate amounts of titanium $^{(IV)}$ isopropoxide and zirconium n-propoxide were added to isopropanol. After 1 hour of vigorous stirring, distilled water was added into the solution dropwise to produce a gel designated ZTO. This was washed several times and dried at 80 °C for 12 hours to obtain ZTO powder. Next, a stoichiometric quantity of ZTO powder was dispersed in 100 mL of NaOH (10 M) solution for 1 hour. Meanwhile, calcium nitrate tetrahydrate and barium acetate were dissolved in 50 mL of distilled water. Then, the two solutions were mixed together in a 250-ml round-bottom flask equipped with a magnetic stirrer for 3 hours under nitrogen flow. After, the suspension obtained was transferred to a Teflon-lined stainless-steel autoclave, purged with nitrogen, sealed and heated at 25, 40, 80, 120, 160 and 240 °C for 24 h and designated B-25, B-40, B-80, B-120, B-160 and B-240, respectively. After the reaction was completed, the sealed autoclave was cooled down to room temperature. The resulting white precipitates were collected by centrifugation at 12 000 rpm for 10 min, and washed several times with distilled water and ethanol. Then, the final products were dried at 100 °C for 12 hours.

*2.2. Characterization*

Crystalline structure of BCZT powder was performed by X-ray diffraction (XRD, Panalytical X-Pert Pro) using a Cu-K$_α$ radiation (λ ~ 1.540598 Å). The resulting microstructures were analyzed using a Scanning Electron Microscope (SEM, Tescan VEGA-3). Fourier transform infrared spectroscopy (FTIR, Bruker VERTEX 70) was used to have an insight on the purity of the synthesized samples, using the KBr in the range of 4000 - 400 cm$^{-1}$.

## 3. Results and discussion

Fig. 1 reveals the XRD patterns of BCZT powders elaborated at different temperatures. It was observed that all BCZT powders possess a pure perovskite phase, without any secondary phase peaks including B-25 elaborated at room temperature. The peaks-splitting similar to those reported by G.K. Sahoo and R. Mazumder[19] were observed at $2\theta \approx 45°$ (Fig. 1(b)) and 65.5° (Fig. 1(c)). These indicate that all BCZT powders were formed at the Morphotropic Phase Boundary (MPB) with the coexistence of orthorhombic and tetragonal phases[19–22]. Meanwhile, the diffraction peaks of all samples shift to higher $2\theta$ values by increasing the hydrothermal temperature. These suggest a decreasing of BCZT unit cell, and/or the reduction of defects and structural relaxation [23,24]. Moreover, the peaks broadening observed after increasing the reaction temperature could be attributed to strain on the cell parameters issued from the cationic disorder (substitution of $Ba^{2+}$ by $Ca^{2+}$)[25].

To have an insight on the purity of the elaborated powders, FTIR measurements were performed. Fig. 2 depicts FTIR spectra of BCZT powders synthesized at different temperatures. The absorption peaks between 850 and 1230 $cm^{-1}$ observed in B-25 were attributed to carbonates compounds. A broad absorption band appeared at about 3400 $cm^{-1}$ was assigned to stretching vibrations of M–OH groups (M: Ba, Ca, Zr, and Ti). It is worth noting that an obvious attenuation of the peaks corresponding to carbonate ($CO_3^{2-}$), hydroxyl (OH) and carboxylate (COOH) groups was clearly seen after increasing the temperature. Whilst, an increase in the intensity of the absorption M–O band located between 500 and 650 $cm^{-1}$ was accompanied to the increase of thermal treatment [26]. These results indicate that pure BCZT can be obtained over 80 °C, and corroborate those obtained in XRD measurements suggesting the enhancement of the crystallinity of samples with the increase of temperature.

Fig. 3 displays the SEM micrographs and particle size distribution histograms ofBCZT powders synthesized at different temperatures. Agglomerated spherical particles were observed in B-40 powder suggesting a growth deficient in this sample. After rising the treatment temperature from 40 °C to 80 °C, mature and spherical shaped BCZT particles were formed with increased particle size from 145 nm to 160 nm respectively. Consequently, combining both sol-gel and hydrothermal processes is a promising route to prepare the pure and high-crystalline BCZT powders with small particle size at very low-temperature (80 °C).

The most plausible mechanisms governing the hydrothermal synthesis of BCZT powders consist in the *in situ* transformation (Fig. 4 (a))and/or dissolution-precipitation processes(Fig. 4 (b))[27,28]. The first involves a mutual reaction between ZTOand Ba/Ca ions to form BCZT layer for barium and calcium diffusion until ZTO particles have completely reacted. However, the second one results from a dissolution-precipitation mechanism involving the break of Ti–O and Zr–O bonds via hydrolytic attackto form titanium and zirconium complexes ($Me(OH)_x^{4-x}$), Me: Ti/Zr)that easily react with barium and calcium ions to form homogeneous and/or heterogeneousBCZTnucleations, witch agglomerate to create BCZT particles.

## 4. Conclusion

A novel strategy to elaborate pure and crystalline BCZT ferroelectric powders at very low temperature (80 °C) was achieved successfully. This approach was based on two simple steps, starting from the preparation of ZTO powder via sol-gel method, followed by a hydrothermal reaction between ZTO and barium/calcium ions at different temperatures in highly alkaline medium. It was clearly shown that pure and crystalline BCZT can be obtained over 80°C and that temperature preparation is a key lever to control the crystallinity, the purity and both the morphology and the mean size of lead-free BCZT powders.These eco-friendly powders can be used without any further treatment in several applications such as biomedical, capacitors, actuators and electromechanical transducers.

**Acknowledgements**

The authors gratefully acknowledge the generous financial support of CNRST Priority Program PPR 15/2015 and H2020-MSCA-RISE-2017-ENGIMA action.

**Figure captions**

**Fig. 1.** (a) XRD patterns of BCZT powders synthesized from 25 to 240 °C, and enlarged peaks splitting at 2θ ≈ 45° (b) and 65.5° (c).

**Fig. 2.** FTIR spectra of BCZT powders synthesized from 25 to 240 °C.

**Fig. 3.** SEM micrographs and particles size distribution histograms of BCZT powders synthesized at different temperatures.

**Fig. 4.** Schematic representation of the mechanisms of BCZT particles formation under hydrothermal processing: (a) *in situ* transformation and (b) dissolution-precipitation processes.

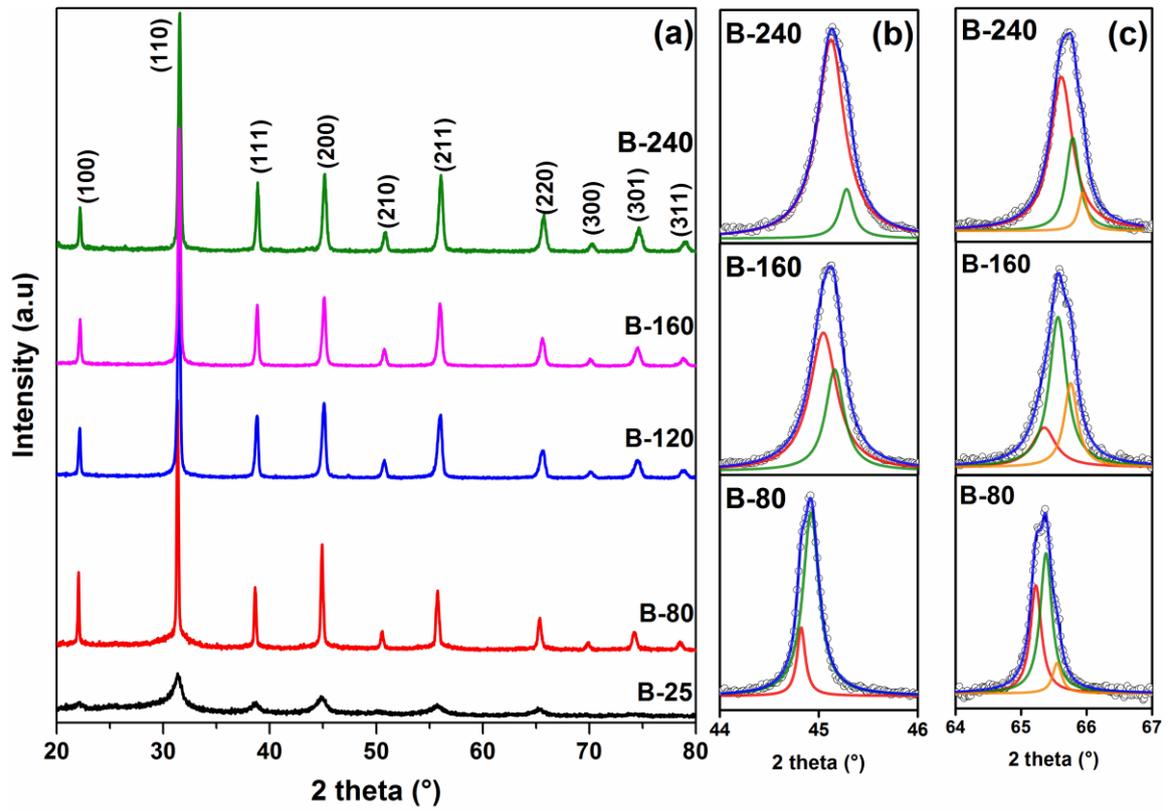

**Fig. 1.** (a) XRD patterns of BCZT powders synthesized from 25 to 240 °C, and enlarged peaks splitting at 2θ ≈ 45° (b) and 65.5° (c).

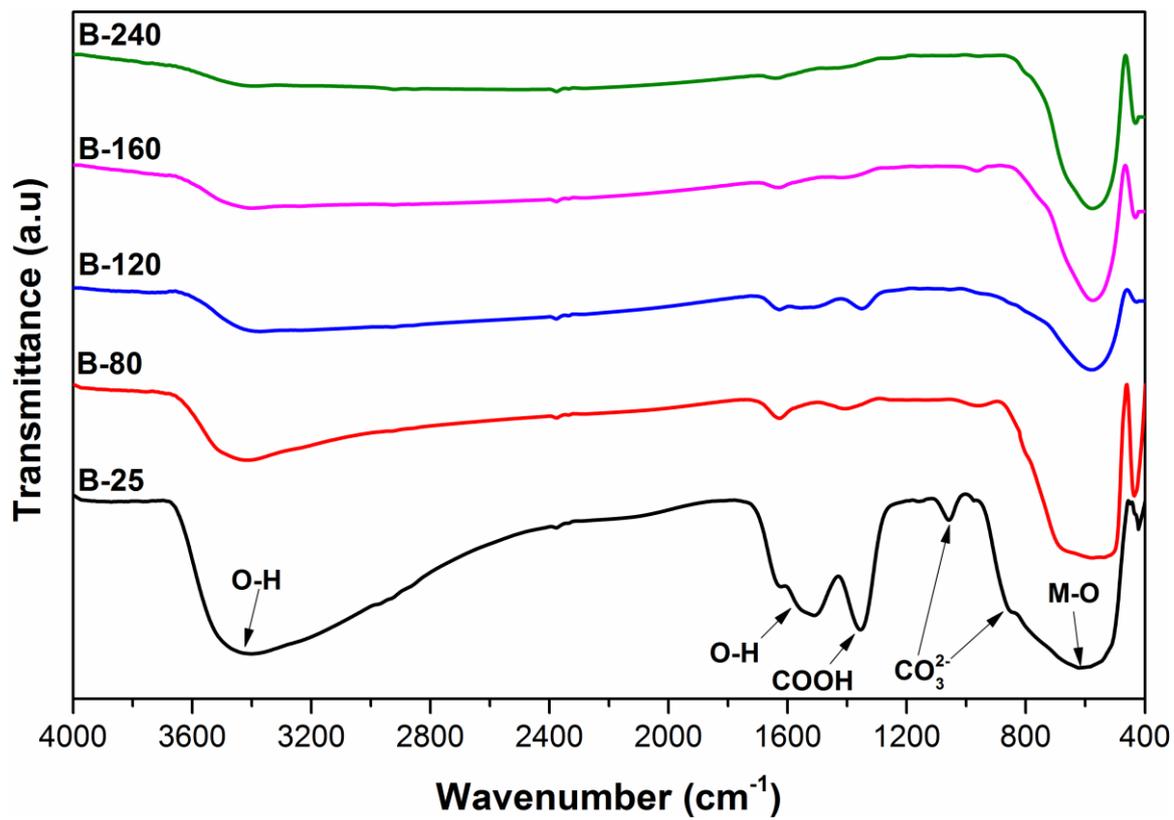

**Fig. 2.** FTIR spectra of BCZT powders synthesized from 25 to 240 °C.

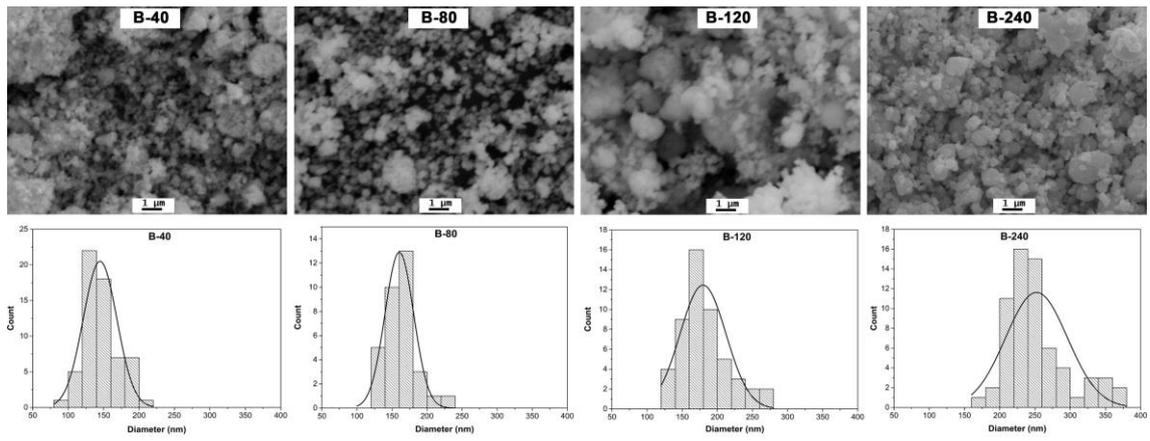

**Fig. 3.** SEM micrographs and particles size distribution histograms of BCZT powders synthesized at different temperatures.

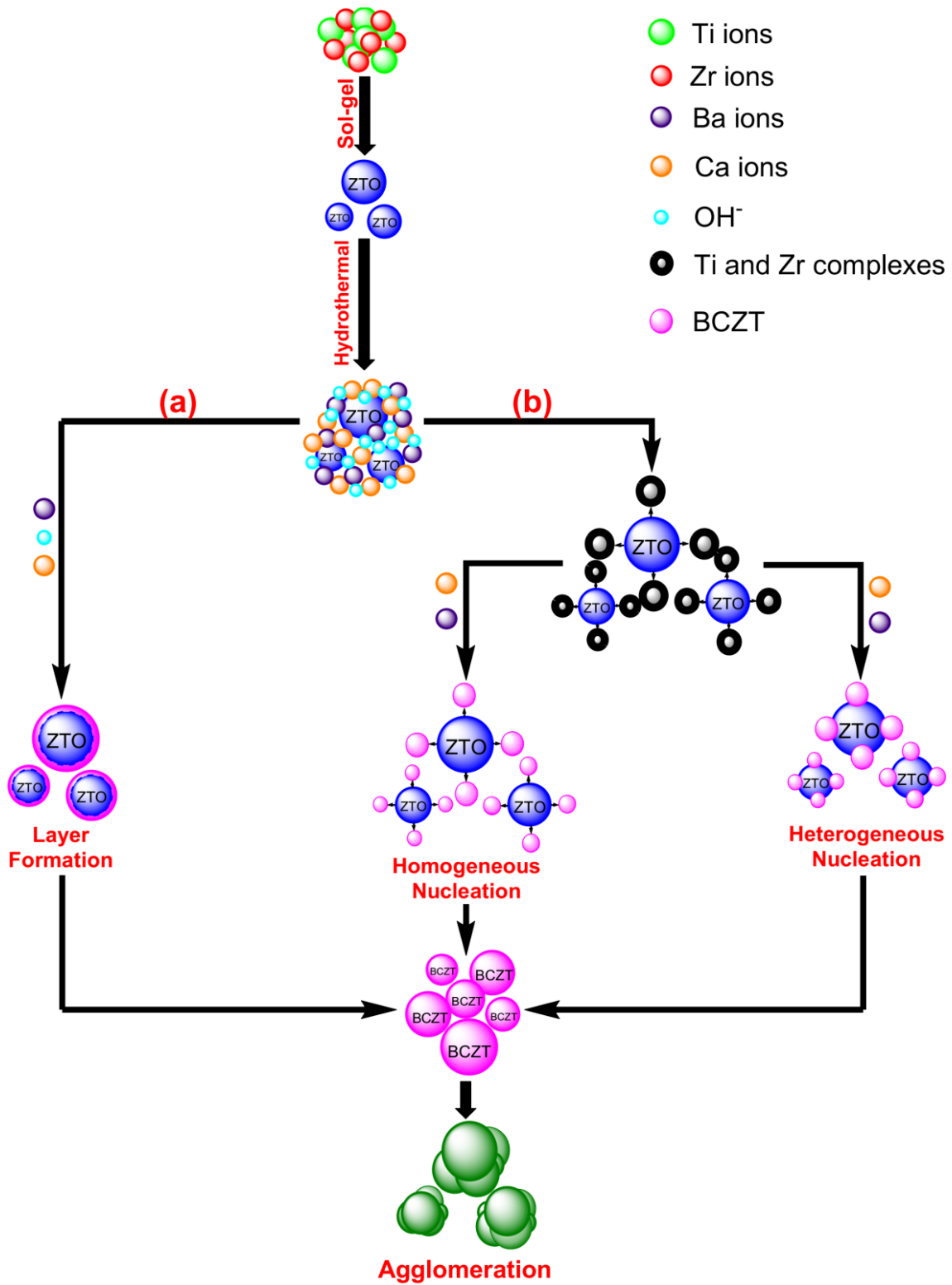

**Fig. 4.** Schematic representation of the mechanisms of BCZT particles formation under hydrothermal processing: (a) *in situ* transformation and (b) dissolution-precipitation processes.